\documentclass[11pt]{article}

\usepackage[preprint]{acl}

\usepackage{times}
\usepackage{latexsym}

\usepackage[T1]{fontenc}

\usepackage[utf8]{inputenc}

\usepackage{microtype}

\usepackage{inconsolata}

\usepackage{graphicx}
\usepackage{booktabs}
\usepackage{tabularx}
\usepackage{tikz}
\usetikzlibrary{arrows.meta}

\newcommand{\ours}{\textsc{Audio-Mind}}

\title{Audio-Mind: An Auditable Agentic Framework for Audio Understanding}

\author{
  \textbf{Yucheng Wang\textsuperscript{1,2\textdagger}} \quad
  \textbf{Jing Peng\textsuperscript{3\textdagger}} \quad
  \textbf{Hanqi Li\textsuperscript{3}} \quad
  \textbf{Chenghao Wang\textsuperscript{4}} \\
  \textbf{Wenming Tu\textsuperscript{3}} \quad
  \textbf{Yu Xi\textsuperscript{3}} \quad
  \textbf{Zhaokai Sun\textsuperscript{5}} \quad
  \textbf{Kai Yu\textsuperscript{3}} \quad
  \textbf{Shuai Wang\textsuperscript{1\textdaggerdbl}} \\
  \\[0.6ex]
  \normalfont\small
  \textsuperscript{1}School of Intelligence Science and Technology, Nanjing University, China \\
  \normalfont\small
  \textsuperscript{2}Department of Computer Science, ETH Zürich, Switzerland \\
  \normalfont\small
  \textsuperscript{3}X-LANCE Lab, School of Computer Science, Shanghai Jiao Tong University, China \\
  \normalfont\small
  \textsuperscript{4}School of Automation Science and Engineering, Xi'an Jiaotong University, China \\
  \normalfont\small
  \textsuperscript{5}School of Computer Science, Northwestern Polytechnical University, China
}

\begin{document}
\maketitle
\begingroup
\renewcommand{\thefootnote}{\textdagger}
\footnotetext{Equal contribution.}
\renewcommand{\thefootnote}{\textdaggerdbl}
\footnotetext{Corresponding author.}

\endgroup
\begin{abstract}
Audio agents extend large audio-language models (LALMs) by decomposing audio questions into tool calls, intermediate evidence, and iterative reasoning steps. However, as LALMs become stronger, the key challenge shifts from enabling tool use to determining when agentic evidence acquisition genuinely benefits audio understanding. We propose \ours{}, an auditable and pluggable framework for conditional evidence acquisition in audio understanding. \ours{} dynamically combines a strong frontend with planner-guided tool use, preserving frontend judgment when initial evidence is sufficient while acquiring bounded external evidence for questions with unresolved evidence gaps. Experiments on MMAR and MSU-Bench show that \ours{} outperforms prior audio-agent baselines, reaching 80.4\% accuracy on MMAR and 82.8\% accuracy on MSU-Bench. A matched-backbone comparison highlights why this design matters: under strong audio frontends, agentic decomposition can become an orchestration bottleneck when the workflow does not preserve the frontend’s holistic audio-grounded judgment. Beyond accuracy, \ours{} produces higher-quality, auditable reasoning traces that expose uncertainty, tool evidence, and answer rationales, offering a potential basis for more reliable audio-QA annotation and error analysis.\footnote{Our \href{https://github.com/DELTA-DoubleWise/Audio-Mind}{code} is publicly available.}
\end{abstract}

\section{Introduction}

Large audio-language models (LALMs) are moving audio understanding beyond task-specific recognition pipelines. Recent systems increasingly accept speech, environmental sounds, and music as native inputs to language models, allowing users to ask natural-language questions over diverse acoustic scenes~\citep{peng2024survey_speech_llms_understanding,cui2024recent_speechlm_survey}. Recent LALM and omni-audio model families have expanded this capability across speech recognition, audio captioning, speaker and scene understanding, music analysis, and audio-grounded dialogue~\citep{chu2024qwen2audio,ding2025kimi,zhang2025mimo,huang2025step,xu2025qwen3,qwen2026qwen35omni}. As these models become stronger, they provide increasingly capable frontend perception and contextual reasoning for audio understanding.

This progress has also encouraged researchers to introduce agentic mechanisms into audio understanding. Existing systems use LLM program generation to construct speech-processing toolchains, coordinate audio models or external tools for multi-step reasoning, transform audio reasoning into iterative textual evidence refinement, or learn policies for when to consult external perception \citep{kuan2024speechcopilot,lee2025audiomaestro,wijngaard2025audiotoolagent,rong2025audiogenie,chen2026audiorouter,tong2026autagent,wan2026speechhands,xiong2025thinkingwithsound,shao2026listeningwithtime}. Together, these studies establish audio agents as an emerging research direction for extending LALMs beyond single-pass audio question answering.

However, much of the current audio-agent literature rests on a premise that becomes less secure as LALMs improve: agentic decomposition is often expected to add useful evidence beyond direct LALM inference. This pressure is already visible in recent Qwen-Omni models which report strong performance not only on general audio understanding benchmarks, but also on ASR, speech translation, and music understanding tasks that have traditionally been served by specialized audio systems~\citep{xu2025qwen3,qwen2026qwen35omni}. \textbf{Under such a strong frontend, the value of external tools depends less on their existence than on whether their outputs are more reliable, more precise, or more targeted than the model's own perception.} This issue is even sharper because many audio properties, such as tempo or chord labels, are inherently ambiguous, convention-dependent, or only partially captured by tool outputs \citep{chiu2022analysis,koops2017chord}. Audio-agent systems therefore need to reason more carefully about when and how to leverage tools, rather than treating tool invocation itself as the source of improvement.

This observation motivates a shift from tool routing to conditional evidence acquisition. In this view, tools are not selected merely because they match a task name; they are invoked only when their outputs can provide bounded, task-relevant evidence that complements the frontend model. A prerequisite for such an agent is therefore a planner-facing description of what each external operation is allowed to contribute: whether it directly provides an audio observation, creates a derived artifact for later inspection, or only validates input assumptions. Making these roles explicit allows the planner to distinguish evidence from preparation, measurements from interpretations, and reliable tool outputs from signals that should remain uncertain.

To operationalize this view, we propose \ours{}, an auditable and pluggable framework for conditional evidence acquisition in audio understanding. \ours{} preserves an audio-capable LALM frontend as the main source of audio-grounded judgment, while a text-LLM planner decides whether bounded tools or targeted re-listening are needed to address a concrete evidence gap. The resulting evidence state records frontend observations, planner decisions, tool outputs, re-listening results, and final rationales, making the answer path auditable rather than treating agentic decomposition as an automatic improvement. This structure also enables higher-quality reasoning traces: by exposing uncertainty, evidence provenance, and intermediate decisions, \ours{} produces explanations that are easier to inspect, compare, and use for audio-QA error analysis.

Our contributions are summarized as follows:

\begin{itemize}
    \item We revisit tool-augmented audio-agent reasoning under strong LALMs, showing that agentic orchestration should be treated as conditional evidence acquisition rather than an automatic improvement over end-to-end LALM inference.

    \item We introduce \ours{}, an auditable and pluggable agentic audio understanding framework with a reliability-aware audio-tool taxonomy and boundary design. The framework separates frontend perception, planning, tool execution, targeted re-listening, evidence fusion, and answer validation in an explicit evidence state.

    \item We evaluate \ours{} on MMAR and MSU-Bench, showing both accuracy gains and improved reasoning traces. \ours{} outperforms prior audio-agent baselines, while matched-backbone comparisons show that existing agentic workflows can become orchestration bottlenecks when they fail to preserve strong frontend perception. Beyond final-answer accuracy, \ours{} produces more auditable and higher-quality reasoning traces for audio-QA inspection and error analysis.
\end{itemize}
\section{Related Work}
\label{sec:related_work}

\subsection{Large Audio-Language Models and Audio Understanding Benchmarks}

Recent work on speech and audio language models has moved from task-specific pipelines toward general-purpose models that accept audio as a native input modality \citep{peng2024survey_speech_llms_understanding,cui2024recent_speechlm_survey}. Models such as SALMONN~\citep{tang2023salmonn}, Qwen2-Audio~\citep{chu2024qwen2audio}, Kimi-Audio~\citep{ding2025kimi}, MiMo-Audio~\citep{zhang2025mimo}, Step-Audio~\citep{huang2025step}, and Audio Flamingo~\citep{kong2024audioflamingo,goel2025audioflamingo3} extend language-model interfaces to speech, sounds, music, and audio-grounded dialogue. More recent omni or reasoning-oriented systems, including Qwen3-Omni~\citep{xu2025qwen3}, Qwen3.5-Omni~\citep{qwen2026qwen35omni}, and Step-Audio-R1~\citep{stepfun2025stepaudior1}, further strengthen audio perception, interaction, and modality-grounded reasoning. These models define increasingly strong frontend baselines for evaluating whether audio-agent mechanisms add useful evidence beyond LALM inference.

Evaluation has likewise moved beyond isolated recognition tasks. Recent benchmarks, including AIR-Bench~\citep{yang2024airbench}, MMAU~\citep{sakshi2024mmau}, MMAR~\citep{ma2025mmar}, MMAU-Pro~\citep{kumar2025mmaupro}, and MSU-Bench~\citep{wang2025msubench}, test increasingly heterogeneous forms of audio understanding, ranging from instruction following and audio reasoning to long-form, multi-audio, spatial, open-ended, and multi-talker scenarios. These benchmarks make strong LALM baselines essential, while also motivating systems that can expose intermediate evidence across diverse audio tasks.

\subsection{Tool-Augmented and Agentic Audio Understanding}

Building on general tool-using LLM agents that interleave reasoning with external actions \citep{yao2022react}, recent audio-agent systems adapt this paradigm to audio understanding. These systems explore different ways to integrate these tools with LALMs. Speech-Copilot~\citep{kuan2024speechcopilot} decomposes speech-processing instructions and generates programs over modular tools, while AudioToolAgent~\citep{wijngaard2025audiotoolagent} coordinates audio-language models and tool adapters through a central LLM agent. Audio-Maestro~\citep{lee2025audiomaestro} and AudioGenie-Reasoner~\citep{rong2025audiogenie} emphasize tool-augmented or multi-agent evidence construction for audio-language reasoning. Other work studies tool-use policy itself: AudioRouter~\citep{chen2026audiorouter} and AuTAgent~\citep{tong2026autagent} learn when and how to invoke tools through reinforcement learning, Speech-Hands~\citep{wan2026speechhands} frames external audio perception as a self-reflection decision, and Multi-Source Evidence Fusion~\citep{olev2026multisource} combines independent LALM observations with acoustic tools organized by reliability tiers. These works establish the promise of agentic audio understanding. However, they primarily study how to introduce or optimize external evidence use, while our work asks when such evidence should be acquired under strong audio frontends and treat tool use as conditional evidence acquisition rather than an automatic improvement over direct LALM inference.

Adjacent systems also show that intermediate audio operations can support understanding even when they are not framed as general-purpose tool agents. Thinking with Sound~\citep{xiong2025thinkingwithsound} combines linguistic reasoning with on-the-fly audio-domain analysis and manipulation, while Listening with Time~\citep{shao2026listeningwithtime} uses global-to-local temporal reasoning with local audio evidence for long-form understanding. Relative to these systems, our framework makes the roles of such intermediate operations more explicit by distinguishing direct evidence extraction from audio transformation, and by exposing both through capability-bounded interfaces.

\begin{figure*}[t]
  \centering
  \includegraphics[width=\textwidth]{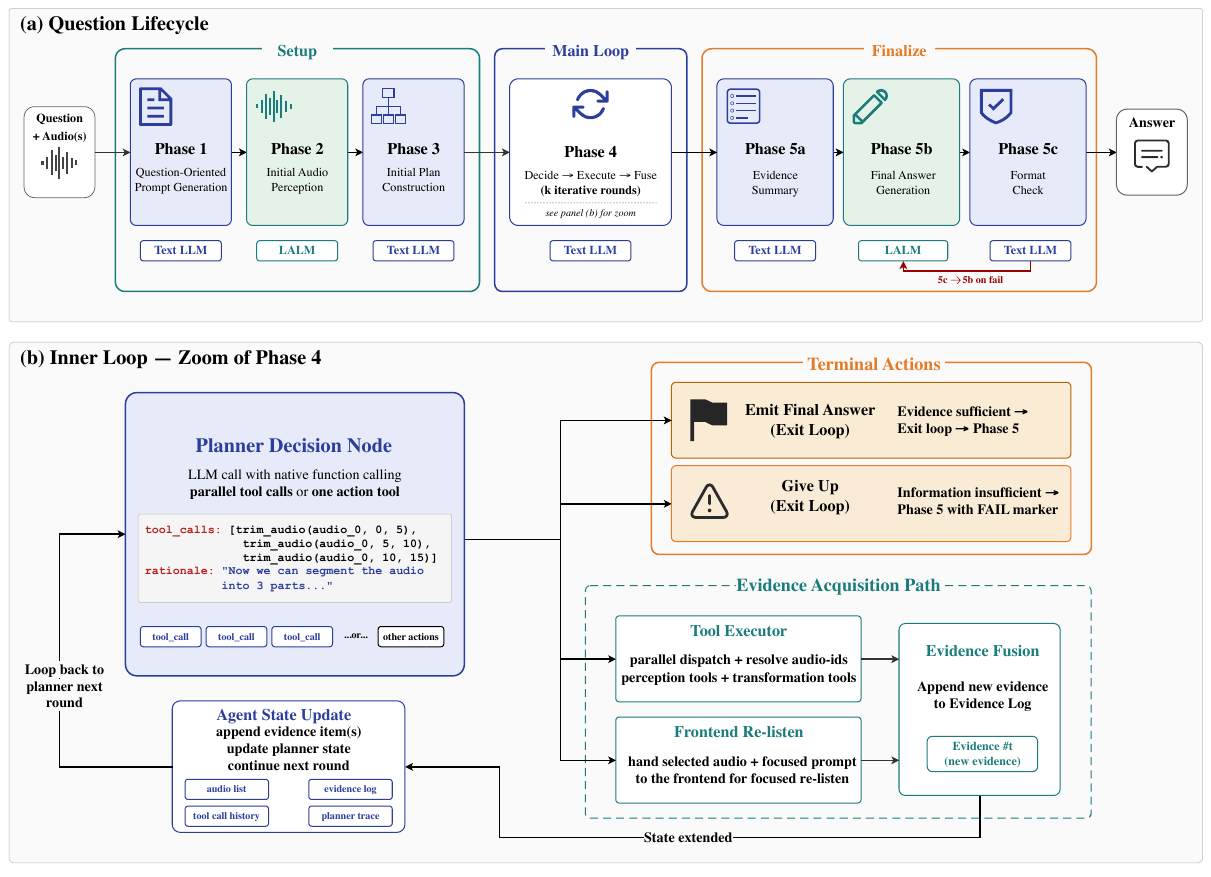}
  \caption{Architecture of \ours{}. The system first uses a planner-generated perception prompt to obtain question-oriented frontend evidence, then iteratively gathers additional evidence through tool calls or targeted frontend follow-up. Once the planner determines that the evidence is sufficient, the frontend model generates the final answer from the original audio and accumulated context, followed by format validation.}
  \label{fig:architecture}
\end{figure*}

\section{Tools as Bounded Evidence Interfaces}
\label{sec:audio_tools}

Before presenting the full agent architecture, we first characterize the role of tool use in audio-agent reasoning. Audio tools play a different role in audio agents than tools usually play in language-only agents. Rather than only expanding the model's external action space, they often operate on the input signal itself: exposing a partial observation of the audio or reshaping the audio so that a later model or tool can inspect it more effectively. Therefore, we treat audio tools as evidence interfaces rather than unconditional substitutes for the frontend LALM.

\subsection{Perception and Transformation Roles}
\label{sec:audio_tools_as_evidence}

Under this view, we distinguish two broad roles. \emph{Perception tools} extract observations from audio, such as transcripts, speaker turns, temporal boundaries, pitch, key, chords, or spectral measurements. These tools can support an answer when the question depends on a local, symbolic, or measurable cue. \emph{Transformation tools} modify the audio source itself, for example by trimming, denoising, separating, filtering, resampling, or converting channels. Their output is not usually evidence by itself; rather, it creates a better input for later perception by a tool or by the frontend model during targeted re-listening.

This distinction matters because not every tool output should be interpreted in the same way. A transcript, segment boundary, acoustic measurement, or derived audio clip may provide strong support for some questions, but may become misleading if the planner treats it as evidence for a broader judgment than it can justify.

\subsection{Tool Boundaries}
\label{sec:tool_boundaries}

Tool boundaries specify what kind of claim a tool output is allowed to support. This is necessary because the same low-level observation can be useful for one question and misleading for another. For example, a transcript can verify lexical content but not speaker identity, while a tempo, chord, or spectral estimate can provide measurable acoustic evidence but not a complete music-theoretic or perceptual judgment. Without such boundaries, an agent may convert a narrow measurement into an unsupported semantic conclusion simply because the output appears precise.

These constraints should not be read as an argument against specialized tools. Rather, boundaries make tool evidence usable by specifying the conditions under which it should be trusted. Even when strong LALMs achieve higher aggregate scores on some audio benchmarks, a frontend LALM is optimized for broad audio-language behavior, while many tools are optimized for narrower interfaces and target scenes with explicit outputs. ASR and diarization systems, for example, may remain more dependable in matched settings such as meetings, interviews, or multi-speaker recordings, where their training objectives and output schemas directly target the needed evidence~\citep{shi2026qwen3asr,xu2025fireredasr,han2025efficient_diarization,park2025sortformer}. These outputs can be easier to inspect, compare, reuse, and pass to later stages than a free-form model response. We therefore treat tool use as conditional evidence acquisition: a tool is valuable when its bounded output provides verifiable support that the frontend either lacks, states uncertainly, or cannot express in the required structure.

\begin{table*}[t]
\small
\centering
\renewcommand{\arraystretch}{1.08}
\setlength{\heavyrulewidth}{0.1em}
\setlength{\lightrulewidth}{0.055em}
\setlength{\cmidrulewidth}{0.035em}
\begin{tabularx}{\textwidth}{>{\raggedright\arraybackslash}p{0.20\textwidth}>{\raggedright\arraybackslash}X>{\raggedright\arraybackslash}p{0.28\textwidth}}
\toprule
\textbf{Category} & \textbf{Description} & \textbf{Representative tools} \\
\midrule
\multicolumn{3}{c}{\textbf{Transformation tools}} \\
\cmidrule(lr){1-3}
Audio derivation (8) & Creates derived audio by trimming, denoising, resampling, conversion, or separation to prepare better inputs for later tools or re-listening. & Time-range trimming; denoising; channel conversion \\
\cmidrule(lr){1-3}
Temporal segmentation (5) & Locates candidate speech, silence, event, or music regions when relevant time spans or boundaries need verification. & Audio segmentation; silence detection; VAD \\
\midrule
\multicolumn{3}{c}{\textbf{Perception tools}} \\
\cmidrule(lr){1-3}
Metadata validation (2) & Checks file-level duration, sampling, channel, and format facts to validate input assumptions rather than audio content. & Audio stream statistics; loading metadata \\
\cmidrule(lr){1-3}
Speech and speaker processing (9) & Verifies speech content, speaker counts, turns, or identity using ASR, diarization, VAD, and speaker verification. & ASR; diarization; speaker verification \\
\cmidrule(lr){1-3}
Acoustic/music feature analysis (12) & Provides supporting measurements such as pitch, spectrum, tempo, key, or chords. & Pitch analysis; tempo estimation; chord recognition \\
\cmidrule(lr){1-3}
Signal visualization inspection (2) & Inspects generated waveforms, spectrograms, or processing artifacts for bounded signal-level cues. & Audio-plot inspection; quality verification \\
\bottomrule
\end{tabularx}
\caption{Planner-facing tool categories used by \ours{}. Numbers in parentheses indicate the number of planner-visible core tools in each category. }
\label{tab:tool_categories}
\end{table*}

\section{Framework Architecture}
\label{sec:method}

\ours{} operationalizes conditional evidence acquisition with three design goals. First, it should preserve the frontend LALM when direct audio perception is already sufficient. Second, it should expose the unresolved evidence gap before invoking tools. Third, it should make each external operation bounded and inspectable, so that tool outputs, transformed audio artifacts, and re-listening results can be traced back to the decision that requested them.

Figure~\ref{fig:architecture} illustrates the architecture of \ours{}. The system is organized as a stateful graph in which each module reads a shared evidence state, performs one bounded operation, and writes structured updates for later modules.

\subsection{System Overview}
\label{sec:system_overview}

Given audio inputs $\mathcal{A}$ and a natural-language question $q$, the system aims to produce an answer $y$ that is grounded in the audio and satisfies the expected output format. The agent is organized around an explicit evidence state that records audio artifacts, frontend observations, planner decisions, tool outputs, evidence summaries, and format feedback. Each graph node reads this state, performs one bounded operation, and writes structured updates for later stages.

In our default instantiation, \ours{} separates audio perception from textual orchestration. An audio-capable LALM serves as the frontend for initial perception, targeted re-listening, and final answer generation, while a text LLM serves as the planner for prompt construction, action selection, evidence summarization, and format checking. The shared evidence state keeps these roles pluggable: frontends, planners, tool executors, and checkers can be replaced as long as they preserve the same evidence contracts.

\subsection{Question-Oriented Perception and Planning}
\label{sec:perception_and_planning}

The first stage is designed to avoid both generic captioning and premature tool use. Directly prompting the frontend with the original question can produce an overcommitted answer or omit the evidence needed for later verification. \ours{} instead asks the planner to generate a self-contained perception prompt that restates the question, names task-specific listening targets, and asks the frontend to report uncertainty. The frontend analyzes $\mathcal{A}$ under this prompt and returns structured evidence: a general description, relevant focus points, a preliminary answer when available, and uncertainty or verification needs.

The planner treats the frontend caption as evidence rather than ground truth. It converts the caption into a lightweight plan that records the clarified intent, expected answer format, focus points, and candidate evidence operations. The plan does not execute tools; it only marks which claims appear answerable by direct perception and which may require verification.

Both the caption and the plan are stored in the shared state. Each input audio file receives a stable identifier such as \texttt{audio\_0}, and any derived artifact created later is appended to the same audio list with its own identifier and provenance. This lets later decisions refer to original or transformed audio consistently while keeping the evidence path inspectable.

\subsection{Evidence Acquisition Loop}
\label{sec:evidence_loop}

The evidence acquisition loop is the point where the agent decides whether additional evidence is worth obtaining. At each iteration, the planner observes the current evidence log, tool-call history, planner trace, available audio artifacts, and available tool inventory. It then emits exactly one action type: call one or more independent tools, request frontend follow-up on selected audio, proceed to final answer generation, or stop with an explicit failure. Each decision must include a short rationale, which is stored in the planner trace and shown back to the planner in later rounds. This makes tool use accountable to a stated evidence gap rather than to tool availability alone.

External tools are exposed through the category-bounded inventory in Table~\ref{tab:tool_categories}. The inventory turns the tool-boundary discussion in Section~\ref{sec:audio_tools} into planner-facing guidance, so action selection depends on what evidence a tool can support rather than on tool names alone. Tool execution returns either an evidence item, a derived audio artifact, or both. This keeps analysis results and transformed inputs in the same shared state, allowing later decisions to use tool outputs without exposing low-level execution details.

Frontend follow-up, or re-listening, is used when the unresolved question is better answered by direct perception on a better-selected input. This is especially useful after tools create a materially improved artifact, such as a trimmed segment containing the relevant event, or a separated speaker track that makes the target source easier to hear. The planner sends selected audio identifiers and a focused prompt to the frontend, and the resulting caption is appended as follow-up evidence. The system does not use re-listening as a generic fallback; it is reserved for cases where selection or transformation changes what the frontend can perceive.

\begin{table*}[t]
\small
\centering
\setlength{\tabcolsep}{4pt}
\resizebox{\textwidth}{!}{%
\begin{tabular}{lcccccccc}
\toprule
\textbf{Method} & \textbf{Sound} & \textbf{Music} & \textbf{Speech} & \textbf{So-Mu} & \textbf{So-Sp} & \textbf{Mu-Sp} & \textbf{So-Mu-Sp} & \textbf{Avg.} \\
\midrule
\multicolumn{9}{c}{\emph{End-to-end models}} \\
\midrule
GPT-4o Audio~\citep{hurst2024gpt} & 53.94 & 50.97 & 70.41 & 63.64 & 72.48 & 62.20 & 75.00 & 63.50 \\
Gemini 2.0 Flash~\citep{google2025gemini20flash} & 61.21 & 50.97 & 72.11 & 81.82 & 72.48 & 65.85 & 70.83 & 65.60 \\
Gemini 2.5 Pro~\citep{comanici2025gemini} & 67.30 & 56.80 & 82.00 & \textbf{100.00} & 84.90 & 80.50 & 66.70 & 74.70 \\
Gemini 3 Pro~\citep{google2026gemini3} & 72.73 & 61.65 & 84.35 & 72.73 & 81.19 & 78.05 & 83.33 & 76.40 \\
Qwen3.5-Omni~\citep{qwen2026qwen35omni} & -- & -- & -- & -- & -- & -- & -- & 80.00 \\
Qwen3.5-Omni$^\dagger$~\citep{qwen2026qwen35omni} & 75.76 & 61.17 & 85.37 & 90.91 & \textbf{87.61} & 79.27 & \textbf{87.50} & 78.90 \\
\midrule
\multicolumn{9}{c}{\emph{Audio agents}} \\
\midrule
AudioRouter~\citep{chen2026audiorouter} & 61.40 & \textbf{66.10} & 77.20 & 57.20 & 65.10 & 68.30 & 55.50 & 64.40 \\
AuTAgent~\citep{tong2026autagent} & 51.52 & 45.15 & 73.81 & -- & -- & -- & -- & 63.00 \\
AudioToolAgent~\citep{wijngaard2025audiotoolagent} & 73.33 & 63.59 & 82.99 & 63.64 & 83.03 & \textbf{81.71} & 79.17 & 77.00 \\
AudioGenie-Reasoner~\citep{rong2025audiogenie} & 49.68 & 43.26 & 69.23 & 45.45 & 64.53 & 65.33 & 59.09 & 58.85 \\
AudioGenie-Reasoner$^\dagger$ (Matched Backbone)~\citep{rong2025audiogenie} & 70.91 & 54.85 & 79.59 & 54.55 & 72.48 & 76.83 & 58.33 & 70.50 \\
\ours{}$^\dagger$ & \textbf{78.18} & 63.11 & \textbf{87.41} & \textbf{100.00} & \textbf{87.61} & 79.27 & \textbf{87.50} & \textbf{80.40} \\
\bottomrule
\end{tabular}
}
\caption{MMAR accuracy by modality. So, Mu, and Sp denote sound, music, and speech. Rows marked with $^\dagger$ are evaluated by us on the same 1,000 MMAR examples; other values are taken from the corresponding benchmark or system reports.}
\label{tab:mmar_main}
\end{table*}

\subsection{Answer Generation and Validation}
\label{sec:answer_generation}

The final stage keeps answer synthesis separate from evidence acquisition. Once the planner judges that the evidence is sufficient, it emits an answer action rather than writing the final response itself. The system then prepares a neutral evidence summary that reports observations, preserves apparent contradictions, and avoids deciding which source is correct. The audio-capable frontend receives this summary together with the original audio and task context, so the final answer is still grounded in direct listening while being informed by accumulated evidence.

A separate format checker then validates only whether the proposed answer follows the expected structure. It does not judge content correctness; it simply returns structural feedback when the answer format is invalid. This prevents formatting errors from being conflated with evidence failures.

\begin{table}[t]
\small
\centering
\resizebox{\columnwidth}{!}{%
\begin{tabular}{lc}
\toprule
\textbf{Model} & \textbf{Avg.} \\
\midrule
\multicolumn{2}{c}{\emph{End-to-end models}} \\
\midrule
Kimi-Audio~\citep{ding2025kimi} & 0.460 \\
StepAudio2~\citep{wu2025step} & 0.480 \\
MiMoAudio~\citep{zhang2025mimo} & 0.600 \\
Gemini 2.5 Flash$^\dagger$~\citep{comanici2025gemini} & 0.682 \\
Qwen3.5-Omni$^\dagger$~\citep{qwen2026qwen35omni} & 0.763 \\
Gemini 2.5 Pro (thinking)$^\dagger$~\citep{comanici2025gemini} & 0.819 \\
\midrule
\multicolumn{2}{c}{\emph{Agentic systems}} \\
\midrule
AudioGenie-Reasoner$^\dagger$ (Matched Backbone)~\citep{rong2025audiogenie} & 0.789 \\
\ours{}$^\dagger$ & \textbf{0.828} \\
\bottomrule
\end{tabular}
}
\caption{Average accuracy on MSU-Bench. Rows marked with $^\dagger$ are evaluated by us; unmarked values are reported by MSU-Bench \citep{wang2025msubench}. AudioGenie-Reasoner and \ours{} use the Gemini configuration for MSU-Bench.}
\label{tab:msubench_main}
\end{table}

\section{Results}

\subsection{Experimental Setup}
\label{sec:experimental_setup}

We evaluate \ours{} on two audio understanding benchmarks: MMAR~\citep{ma2025mmar} and MSU-Bench~\citep{wang2025msubench}. MMAR covers a broad range of audio question-answering categories, while MSU-Bench focuses on information-dense multi-speaker understanding.

For each benchmark, we use the strongest accessible LALM as the audio-capable frontend. On MMAR, \ours{} uses Qwen3.5-Omni as the frontend model and Qwen3.5 as the text planner. On MSU-Bench, \ours{} uses Gemini 2.5 Pro as the frontend model and Gemini 3.0 Flash as the text planner. We also evaluate the corresponding LALM baseline, where the audio-capable model answers the benchmark question without agentic tool use.

We compare against AudioGenie-Reasoner as an agentic baseline~\citep{rong2025audiogenie}. Our main purpose is to test how its agentic design behaves when paired with a strong LALM frontend, so we instantiate AudioGenie-Reasoner with the matched backbone for each benchmark: Qwen3.5-Omni and Qwen3.5 on MMAR, and the Gemini configuration on MSU-Bench. All systems are evaluated with the official multiple-choice answer format of each benchmark, and predictions are scored by matching the selected option against the ground-truth answer.

\subsection{Overall Results}
\label{sec:mmar_results}

Tables~\ref{tab:mmar_main} and~\ref{tab:msubench_main} report the main benchmark results. On MMAR, \ours{} reaches 80.40 average accuracy, improving over our Qwen3.5-Omni evaluation at 78.90 while also outperforming the prior audio-agent baselines in the table. On MSU-Bench, \ours{} reaches 0.828 accuracy, slightly above the direct Gemini 2.5 Pro counterpart at 0.819 and the matched-setting AudioGenie-Reasoner result at 0.789. These gains over strong LALM counterparts are modest, but they are consistent with the intended design: the agent preserves strong frontend perception as the final audio-grounded decision point while selectively acquiring external evidence when the planner identifies a concrete uncertainty or verification need. The MMAR sub-category results further suggest where this design is most useful. \ours{} shows larger gains on Music Theory and Temporal Analysis, where structured music measurements and localized or transformed audio segments can provide evidence that is difficult for a single-pass frontend response to retain precisely. Appendix~\ref{sec:mmar_subcategory_appendix} provides the full MMAR sub-category breakdown for Qwen3.5-Omni, matched-backbone AudioGenie-Reasoner, and \ours{}.

The matched-backbone AudioGenie-Reasoner result illustrates why stronger base models do not automatically make an agentic pipeline stronger. On MMAR, AudioGenie-Reasoner instantiated with Qwen3.5-Omni and Qwen3.5 rises substantially over its originally reported result (70.50 vs. 58.85 average accuracy), showing that the stronger backbone helps. However, it still falls noticeably below Qwen3.5-Omni in our evaluation on MMAR (70.50 vs. 78.90). On MSU-Bench, using the same Gemini setting as \ours{}, AudioGenie-Reasoner reaches 0.789, below both Gemini 2.5 Pro direct inference at 0.819 and \ours{} at 0.828. This suggests that some agentic designs can underuse a strong omni frontend: if the audio model is mainly treated as an intermediate evidence generator, and the final decision is routed through compressed textual traces, the pipeline may lose part of the holistic perception and reasoning ability already available in the frontend model. Thus, agentic audio reasoning should not be viewed as a guaranteed improvement over LALM inference; its benefit depends on whether the agent preserves reliable frontend evidence while adding useful external evidence.

\subsection{Behavioral Analysis of Tool Use}
\label{sec:behavior_analysis}

To examine whether \ours{} behaves the way conditional evidence acquisition predicts, we stratify the 999 successfully completed MMAR runs by the number of tool calls the planner makes per question and compare against the Qwen3.5-Omni baseline on the same items. Across all examples the planner uses on average 1.68 tool calls per question, and 31.2\% of questions exit at zero tool calls. The framework is therefore empirically conditional rather than always-on.

Table~\ref{tab:tool_call_stratified_accuracy} reveals two coupled trends. First, Qwen3.5-Omni accuracy decreases monotonically as tool-call count grows, from 90.4\% at zero calls to 52.0\% in the 6--10 bucket. Since the planner has no access to ground-truth labels or baseline outcomes, and only observes the question, frontend-generated evidence, accumulated state, and available tool interfaces, this alignment suggests that \ours{} captures a practical, question-level estimate of the frontend LALM's capability boundary through its evidence-gap formulation: tool-use depth increases when the initial frontend evidence appears insufficient and remains low when it appears adequate. Thus, conditional evidence acquisition is reflected in the agent's behavior, not only in its design. Second, the \ours{}--Qwen3.5-Omni gap becomes positive in higher tool-use buckets. With zero or one tool call, \ours{} remains close to direct inference ($\Delta \in [-0.8,-0.6]$), suggesting that it largely preserves frontend performance when little additional evidence is needed. Gains emerge from two calls onward, reaching +9.4 points in the 4--5 call bucket, indicating that tool-use benefits are concentrated in cases where the planner commits to deeper evidence acquisition rather than uniformly improving all questions.

Together, these trends help explain the aggregate $+1.5$-point improvement: \ours{} remains close to direct frontend inference when the planner requests little or no tool evidence, while its gains concentrate on questions that trigger deeper evidence acquisition, which is consistent with the intended behavior of conditional evidence acquisition. Appendix~\ref{sec:behavior_appendix} reports planner-round and tool-frequency statistics, including the long-tail $>10$-call bucket.

\begin{table}[t]
\small
\centering
\resizebox{\columnwidth}{!}{%
\begin{tabular}{lrrrr}
\toprule
\textbf{Tool calls} & \textbf{N} & \textbf{\ours{}} & \textbf{Qwen3.5-Omni} & \textbf{$\Delta$} \\
\midrule
0 & 311 & 89.7 & 90.4 & -0.6 \\
1 & 255 & 85.5 & 86.3 & -0.8 \\
2 & 164 & 73.8 & 72.6 & +1.2 \\
3 & 142 & 72.5 & 66.2 & +6.3 \\
4--5 & 96 & 69.8 & 60.4 & +9.4 \\
6--10 & 25 & 56.0 & 52.0 & +4.0 \\
$>$10 & 6 & 16.7 & 50.0 & -33.3 \\
\bottomrule
\end{tabular}
}
\caption{MMAR accuracy comparison between \ours{} and Qwen3.5-Omni, grouped by the number of tool calls made by \ours{}. Accuracy and $\Delta$ are reported in percentage points.}
\label{tab:tool_call_stratified_accuracy}
\end{table}

\subsection{Reasoning Trace Quality Beyond Accuracy}
\label{sec:beyond_accuracy}

\ours{} is designed to make the reasoning process more interpretable by construction. A LALM answer usually compresses perception, reasoning, and justification into a single response, making it difficult to tell which audio cues were actually heard, which uncertainties were ignored, and whether the explanation is faithful to the decision. In contrast, \ours{} separates the trace into explicit stages. This structure does not guarantee correctness, but it makes the answer path easier to audit because unsupported assumptions, conflicting evidence, and tool-dependent claims remain visible.

For reasoning-trace quality, we apply MMAR-Rubrics, the official instance-level protocol from the Interspeech 2026 Audio Reasoning Challenge~\citep{ma2026audiochallenge}. It scores each correct-answer trace by averaging five instance-specific binary criteria and assigns zero to incorrect answers. Under this protocol, \ours{} obtains an MMAR-Rubrics score of 66.5\%, compared with 59.6\% for Qwen3.5-Omni.\footnote{Scores are not directly comparable to the challenge report because we use Qwen3.5 as the judge instead of GPT-4o.} This result suggests that \ours{} can provide higher-quality audio-QA reasoning chains, which may be valuable for downstream data annotation, error analysis, and human inspection beyond aggregate benchmark accuracy. 

Appendix~\ref{sec:case_study} provides a qualitative example of how such traces support inspection. The case illustrates three recurring behaviors: an LALM may hallucinate a plausible acoustic rationale, the question-oriented frontend prompt can expose uncertainty, and bounded tools can provide the missing evidence. This example complements the MMAR-Rubrics results by showing how auditable traces make the correction process visible rather than only reporting a final answer.

\section{Conclusion}

We presented \ours{}, an auditable and pluggable framework for audio understanding under strong LALM frontends. Motivated by the observation that agentic decomposition is not automatically beneficial when LALMs become stronger, \ours{} treats tool use as conditional evidence acquisition. It preserves frontend audio-grounded judgment while selectively using bounded tools and targeted re-listening when additional evidence is needed.

Experiments on MMAR and MSU-Bench show that \ours{} outperforms prior audio-agent baselines and produces more auditable reasoning traces. Behavioral analysis further shows that \ours{} turns conditional evidence acquisition into observable tool allocation, with gains concentrated where additional evidence is most needed. These results suggest that the value of audio agents lies not in more tool calls, but in acquiring the right evidence while keeping uncertainty, tool outputs, re-listening decisions, and final rationales inspectable. This makes \ours{} a useful basis for reliable audio-QA reasoning, annotation, and error analysis.

\newpage
\section*{Limitations}

Our evaluation is limited to MMAR and MSU-Bench, which cover broad audio-QA and multi-speaker understanding but do not fully represent real-world uses such as long-form meeting analysis, audio editing, forensic listening, or interactive inspection. \ours{} also depends on the quality of external tools: although tool outputs are treated as bounded evidence rather than ground truth, they may still be noisy, domain-limited, or convention-dependent, especially for music and acoustic analysis. Finally, the framework introduces additional computation through planning, tool execution, and targeted re-listening, so direct LALM inference may remain preferable for simple or latency-sensitive questions.

\section*{Ethical Considerations}

This work uses existing benchmarks, pretrained models, and external audio tools for research evaluation only. We do not collect new human-subject data or recruit human annotators, and we do not redistribute third-party model weights, benchmark data, or tool artifacts. Since audio-agent systems may be used in sensitive speech or speaker-related settings, outputs from \ours{} should be treated as decision support rather than ground truth, especially for personal or legally consequential audio.



\bibliography{custom}

\begin{thebibliography}{50}
\providecommand{\natexlab}[1]{#1}

\bibitem[{Bai et~al.(2025)Bai, Cai, Chen, Chen, Chen, Cheng, Deng, Ding, Gao, Ge et~al.}]{bai2025qwen3}
Shuai Bai, Yuxuan Cai, Ruizhe Chen, Keqin Chen, Xionghui Chen, Zesen Cheng, Lianghao Deng, Wei Ding, Chang Gao, Chunjiang Ge, and 1 others. 2025.
\newblock Qwen3-vl technical report.
\newblock \emph{arXiv preprint arXiv:2511.21631}.

\bibitem[{Bain et~al.(2023)Bain, Huh, Han, and Zisserman}]{bain2023whisperx}
Max Bain, Jaesung Huh, Tengda Han, and Andrew Zisserman. 2023.
\newblock \href {https://doi.org/10.48550/arXiv.2303.00747} {Whisperx: Time-accurate speech transcription of long-form audio}.
\newblock In \emph{Proceedings of Interspeech 2023}.

\bibitem[{Bredin(2023)}]{bredin2023pyannote}
Herv{\'e} Bredin. 2023.
\newblock pyannote.audio 2.1 speaker diarization pipeline: Principle, benchmark, and recipe.
\newblock In \emph{Proceedings of Interspeech 2023}.

\bibitem[{Chen et~al.(2026)Chen, Chen, Cai, Li, Ye, and Wang}]{chen2026audiorouter}
Liyang Chen, Hongkai Chen, Yujun Cai, Sifan Li, Qingwen Ye, and Yiwei Wang. 2026.
\newblock \href {https://doi.org/10.48550/arXiv.2602.10439} {Audiorouter: Data efficient audio understanding via rl based dual reasoning}.
\newblock \emph{arXiv preprint arXiv:2602.10439}.

\bibitem[{Chiu et~al.(2022)Chiu, M{\"u}ller, Davies, Su, and Yang}]{chiu2022analysis}
Ching-Yu Chiu, Meinard M{\"u}ller, Matthew~EP Davies, Alvin Wen-Yu Su, and Yi-Hsuan Yang. 2022.
\newblock An analysis method for metric-level switching in beat tracking.
\newblock \emph{IEEE Signal Processing Letters}, 29:2153--2157.

\bibitem[{Chu et~al.(2024)Chu, Xu, Yang, Wei, Wei, Guo, Leng, Lv, He, Lin, Zhou, and Zhou}]{chu2024qwen2audio}
Yunfei Chu, Jin Xu, Qian Yang, Haojie Wei, Xipin Wei, Zhifang Guo, Yichong Leng, Yuanjun Lv, Jinzheng He, Junyang Lin, Chang Zhou, and Jingren Zhou. 2024.
\newblock \href {https://doi.org/10.48550/arXiv.2407.10759} {Qwen2-audio technical report}.
\newblock \emph{arXiv preprint arXiv:2407.10759}.

\bibitem[{Comanici et~al.(2025)Comanici, Bieber, Schaekermann, Pasupat, Sachdeva, Dhillon, Blistein, Ram, Zhang, Rosen et~al.}]{comanici2025gemini}
Gheorghe Comanici, Eric Bieber, Mike Schaekermann, Ice Pasupat, Noveen Sachdeva, Inderjit Dhillon, Marcel Blistein, Ori Ram, Dan Zhang, Evan Rosen, and 1 others. 2025.
\newblock Gemini 2.5: Pushing the frontier with advanced reasoning, multimodality, long context, and next generation agentic capabilities.
\newblock \emph{arXiv preprint arXiv:2507.06261}.

\bibitem[{Cui et~al.(2024)Cui, Yu, Jiao, Meng, Zhang, Wang, Guo, and King}]{cui2024recent_speechlm_survey}
Wenqian Cui, Dianzhi Yu, Xiaoqi Jiao, Ziqiao Meng, Guangyan Zhang, Qichao Wang, Yiwen Guo, and Irwin King. 2024.
\newblock \href {https://doi.org/10.48550/arXiv.2410.03751} {Recent advances in speech language models: A survey}.
\newblock \emph{arXiv preprint arXiv:2410.03751}.

\bibitem[{Ding et~al.(2025)Ding, Ju, Leng, Liu, Liu, Shang, Shen, Song, Tan, Tang et~al.}]{ding2025kimi}
Ding Ding, Zeqian Ju, Yichong Leng, Songxiang Liu, Tong Liu, Zeyu Shang, Kai Shen, Wei Song, Xu~Tan, Heyi Tang, and 1 others. 2025.
\newblock Kimi-audio technical report.
\newblock \emph{arXiv preprint arXiv:2504.18425}.

\bibitem[{Goel et~al.(2025)Goel, Ghosh, Kim, Kumar, Kong, Lee, Yang, Duraiswami, Manocha, Valle, and Catanzaro}]{goel2025audioflamingo3}
Arushi Goel, Sreyan Ghosh, Jaehyeon Kim, Sonal Kumar, Zhifeng Kong, Sang-gil Lee, Chao-Han~Huck Yang, Ramani Duraiswami, Dinesh Manocha, Rafael Valle, and Bryan Catanzaro. 2025.
\newblock \href {https://doi.org/10.48550/arXiv.2507.08128} {Audio flamingo 3: Advancing audio intelligence with fully open large audio language models}.
\newblock \emph{arXiv preprint arXiv:2507.08128}.

\bibitem[{{Google}(2025)}]{google2025gemini20flash}
{Google}. 2025.
\newblock \href {https://modelcards.withgoogle.com/assets/documents/gemini-2-flash.pdf} {Gemini 2.0 flash model card}.

\bibitem[{{Google DeepMind}(2026)}]{google2026gemini3}
{Google DeepMind}. 2026.
\newblock \href {https://deepmind.google/gemini} {Gemini 3}.

\bibitem[{Han et~al.(2025)Han, P{\'a}lka, Delcroix, Landini, Rohdin, Cernock{\'y}, and Burget}]{han2025efficient_diarization}
Jiangyu Han, Petr P{\'a}lka, Marc Delcroix, Federico Landini, Johan Rohdin, Jan Cernock{\'y}, and Luk{\'a}{\v{s}} Burget. 2025.
\newblock \href {https://doi.org/10.48550/arXiv.2506.18623} {Efficient and generalizable speaker diarization via structured pruning of self-supervised models}.
\newblock \emph{arXiv preprint arXiv:2506.18623}.

\bibitem[{Huang et~al.(2025)Huang, Wu, Wang, Yan, Hu, Feng, Tian, Shen, Li, Chen et~al.}]{huang2025step}
Ailin Huang, Boyong Wu, Bruce Wang, Chao Yan, Chen Hu, Chengli Feng, Fei Tian, Feiyu Shen, Jingbei Li, Mingrui Chen, and 1 others. 2025.
\newblock Step-audio: Unified understanding and generation in intelligent speech interaction.
\newblock \emph{arXiv preprint arXiv:2502.11946}.

\bibitem[{Hurst et~al.(2024)Hurst, Lerer, Goucher, Perelman, Ramesh, Clark, Ostrow, Welihinda, Hayes, Radford et~al.}]{hurst2024gpt}
Aaron Hurst, Adam Lerer, Adam~P Goucher, Adam Perelman, Aditya Ramesh, Aidan Clark, AJ~Ostrow, Akila Welihinda, Alan Hayes, Alec Radford, and 1 others. 2024.
\newblock Gpt-4o system card.
\newblock \emph{arXiv preprint arXiv:2410.21276}.

\bibitem[{Jiang et~al.(2019)Jiang, Chen, Li, and Xia}]{jiang2019largevocabularychord}
Junyan Jiang, Ke~Chen, Wei Li, and Gus Xia. 2019.
\newblock \href {http://archives.ismir.net/ismir2019/paper/000078.pdf} {Large-vocabulary chord transcription via chord structure decomposition}.
\newblock In \emph{Proceedings of the 20th International Society for Music Information Retrieval Conference}, pages 644--651, Delft, The Netherlands.

\bibitem[{Kong et~al.(2024)Kong, Goel, Badlani, Ping, Valle, and Catanzaro}]{kong2024audioflamingo}
Zhifeng Kong, Arushi Goel, Rohan Badlani, Wei Ping, Rafael Valle, and Bryan Catanzaro. 2024.
\newblock \href {https://doi.org/10.48550/arXiv.2402.01831} {Audio flamingo: A novel audio language model with few-shot learning and dialogue abilities}.
\newblock \emph{arXiv preprint arXiv:2402.01831}.

\bibitem[{Koops et~al.(2017)Koops, de~Haas, Bransen, and Volk}]{koops2017chord}
Hendrik~Vincent Koops, W~Bas de~Haas, Jeroen Bransen, and Anja Volk. 2017.
\newblock Chord label personalization through deep learning of integrated harmonic interval-based representations.
\newblock \emph{arXiv preprint arXiv:1706.09552}.

\bibitem[{Kuan et~al.(2024)Kuan, Yang, Huang, Lu, and Lee}]{kuan2024speechcopilot}
Chun-Yi Kuan, Chih-Kai Yang, Wei-Ping Huang, Ke-Han Lu, and Hung-yi Lee. 2024.
\newblock \href {https://doi.org/10.48550/arXiv.2407.09886} {Speech-copilot: Leveraging large language models for speech processing via task decomposition, modularization, and program generation}.
\newblock \emph{arXiv preprint arXiv:2407.09886}.

\bibitem[{Kumar et~al.(2025)Kumar, Sedl{\'a}{\v c}ek, Lokegaonkar, L{\'o}pez, Yu, Anand, Ryu, Chen, Pli{\v c}ka, Hlav{\'a}{\v c}ek et~al.}]{kumar2025mmaupro}
Sonal Kumar, {\v S}imon Sedl{\'a}{\v c}ek, Vaibhavi Lokegaonkar, Fernando L{\'o}pez, Wenyi Yu, Nishit Anand, Hyeonggon Ryu, Lichang Chen, Maxim Pli{\v c}ka, Miroslav Hlav{\'a}{\v c}ek, and 1 others. 2025.
\newblock \href {https://doi.org/10.48550/arXiv.2508.13992} {{MMAU-Pro}: A challenging and comprehensive benchmark for holistic evaluation of audio general intelligence}.
\newblock \emph{arXiv preprint arXiv:2508.13992}.

\bibitem[{Lee et~al.(2025)Lee, Lin, and Lee}]{lee2025audiomaestro}
Kuan-Yi Lee, Tsung-En Lin, and Hung-Yi Lee. 2025.
\newblock \href {https://doi.org/10.48550/arXiv.2510.11454} {Audio-maestro: Enhancing large audio-language models with tool-augmented reasoning}.
\newblock \emph{arXiv preprint arXiv:2510.11454}.

\bibitem[{Ma et~al.(2025)Ma, Ma, Zhu, Yang, Chao, Xu, Chen, Chen, Chen, Cong et~al.}]{ma2025mmar}
Ziyang Ma, Yinghao Ma, Yanqiao Zhu, Chen Yang, Yi-Wen Chao, Ruiyang Xu, Wenxi Chen, Yuanzhe Chen, Zhuo Chen, Jian Cong, and 1 others. 2025.
\newblock \href {https://doi.org/10.48550/arXiv.2505.13032} {{MMAR}: A challenging benchmark for deep reasoning in speech, audio, music, and their mix}.
\newblock \emph{arXiv preprint arXiv:2505.13032}.

\bibitem[{Ma et~al.(2026)Ma, Xu, Ma, Yang, Li, Kim, Xu, Li, Busso, Yu, Chng, and Chen}]{ma2026audiochallenge}
Ziyang Ma, Ruiyang Xu, Yinghao Ma, Chao-Han~Huck Yang, Bohan Li, Jaeyeon Kim, Jin Xu, Jinyu Li, Carlos Busso, Kai Yu, Eng~Siong Chng, and Xie Chen. 2026.
\newblock \href {https://doi.org/10.48550/arXiv.2602.14224} {The interspeech 2026 audio reasoning challenge: Evaluating reasoning process quality for audio reasoning models and agents}.
\newblock \emph{arXiv preprint arXiv:2602.14224}.

\bibitem[{McFee et~al.(2015)McFee, Raffel, Liang, Ellis, McVicar, Battenberg, Nieto et~al.}]{mcfee2015librosa}
Brian McFee, Colin Raffel, Dawen Liang, Daniel~PW Ellis, Matt McVicar, Eric Battenberg, Oriol Nieto, and 1 others. 2015.
\newblock librosa: Audio and music signal analysis in python.
\newblock \emph{SciPy}, 2015(18-24):7.

\bibitem[{Olev and Alum{\"a}e(2026)}]{olev2026multisource}
Aivo Olev and Tanel Alum{\"a}e. 2026.
\newblock \href {https://doi.org/10.48550/arXiv.2603.17822} {Multi-source evidence fusion for audio question answering}.
\newblock \emph{arXiv preprint arXiv:2603.17822}.

\bibitem[{Park et~al.(2025)Park, Medennikov, Dhawan, Wang, Huang, Koluguri, Puvvada, Balam, and Ginsburg}]{park2025sortformer}
Taejin Park, Ivan Medennikov, Kunal Dhawan, Weiqing Wang, He~Huang, Nithin~Rao Koluguri, Krishna~C. Puvvada, Jagadeesh Balam, and Boris Ginsburg. 2025.
\newblock \href {https://doi.org/10.48550/arXiv.2409.06656} {Sortformer: A novel approach for permutation-resolved speaker supervision in speech-to-text systems}.
\newblock In \emph{Proceedings of the 42nd International Conference on Machine Learning}.

\bibitem[{Peng et~al.(2024)Peng, Wang, Li, Guo, Wang, Fang, Xi, Li, Li, Zhang, Wang, and Yu}]{peng2024survey_speech_llms_understanding}
Jing Peng, Yucheng Wang, Bohan Li, Yiwei Guo, Hankun Wang, Yangui Fang, Yu~Xi, Haoyu Li, Xu~Li, Ke~Zhang, Shuai Wang, and Kai Yu. 2024.
\newblock \href {https://doi.org/10.48550/arXiv.2410.18908} {A survey on speech large language models for understanding}.
\newblock \emph{arXiv preprint arXiv:2410.18908}.

\bibitem[{Plaquet and Bredin(2023)}]{plaquet2023powerset}
Alexis Plaquet and Herv{\'e} Bredin. 2023.
\newblock Powerset multi-class cross entropy loss for neural speaker diarization.
\newblock In \emph{Proceedings of Interspeech 2023}.

\bibitem[{{Qwen Team}(2026)}]{qwen2026qwen35omni}
{Qwen Team}. 2026.
\newblock \href {https://doi.org/10.48550/arXiv.2604.15804} {Qwen3.5-omni technical report}.
\newblock \emph{arXiv preprint arXiv:2604.15804}.

\bibitem[{Rong et~al.(2025)Rong, Li, Yu, and Liu}]{rong2025audiogenie}
Yan Rong, Chenxing Li, Dong Yu, and Li~Liu. 2025.
\newblock \href {https://doi.org/10.48550/arXiv.2509.16971} {Audiogenie-reasoner: A training-free multi-agent framework for coarse-to-fine audio deep reasoning}.
\newblock \emph{arXiv preprint arXiv:2509.16971}.

\bibitem[{Sakshi et~al.(2024)Sakshi, Tyagi, Kumar, Seth, Selvakumar, Nieto, Duraiswami, Ghosh, and Manocha}]{sakshi2024mmau}
S.~Sakshi, Utkarsh Tyagi, Sonal Kumar, Ashish Seth, Ramaneswaran Selvakumar, Oriol Nieto, Ramani Duraiswami, Sreyan Ghosh, and Dinesh Manocha. 2024.
\newblock \href {https://doi.org/10.48550/arXiv.2410.19168} {{MMAU}: A massive multi-task audio understanding and reasoning benchmark}.
\newblock \emph{arXiv preprint arXiv:2410.19168}.

\bibitem[{Schreiber and M{\"u}ller(2018)}]{schreiber2018tempocnn}
Hendrik Schreiber and Meinard M{\"u}ller. 2018.
\newblock \href {https://doi.org/10.5281/zenodo.1492353} {A single-step approach to musical tempo estimation using a convolutional neural network}.
\newblock In \emph{Proceedings of the 19th International Society for Music Information Retrieval Conference}, pages 98--105, Paris, France.

\bibitem[{Shao et~al.(2026)Shao, Su, Tian, Mu, Lin, Fan, Luo, Luan, and Xie}]{shao2026listeningwithtime}
Mingchen Shao, Hang Su, Wenjie Tian, Bingshen Mu, Zhennan Lin, Lichun Fan, Zhenbo Luo, Jian Luan, and Lei Xie. 2026.
\newblock \href {https://doi.org/10.48550/arXiv.2604.22245} {Listening with time: Precise temporal awareness for long-form audio understanding}.
\newblock \emph{arXiv preprint arXiv:2604.22245}.

\bibitem[{Shi et~al.(2026)Shi, Wang, Guo, Wang, Zhang, Zhang, Guo, Hao, Xi, Yang, Xu, Zhou, and Lin}]{shi2026qwen3asr}
Xian Shi, Xiong Wang, Zhifang Guo, Yongqi Wang, Pei Zhang, Xinyu Zhang, Zishan Guo, Hongkun Hao, Yu~Xi, Baosong Yang, Jin Xu, Jingren Zhou, and Junyang Lin. 2026.
\newblock \href {https://doi.org/10.48550/arXiv.2601.21337} {Qwen3-asr technical report}.
\newblock \emph{arXiv preprint arXiv:2601.21337}.

\bibitem[{Tang et~al.(2023)Tang, Yu, Sun, Chen, Tan, Li, Lu, MA, and Zhang}]{tang2023salmonn}
Changli Tang, Wenyi Yu, Guangzhi Sun, Xianzhao Chen, Tian Tan, Wei Li, Lu~Lu, Zejun MA, and Chao Zhang. 2023.
\newblock \href {https://doi.org/10.48550/arXiv.2310.13289} {{SALMONN}: Towards generic hearing abilities for large language models}.
\newblock \emph{arXiv preprint arXiv:2310.13289}.

\bibitem[{Tian et~al.(2025)Tian, Zhang, Zhang, Zhang, Li, Liu, Deng, Wu, Chen, Zhao, Yao, Liu, Chng, Yang, Zhang, Jiang, and Yu}]{stepfun2025stepaudior1}
Fei Tian, Xiangyu~Tony Zhang, Yuxin Zhang, Haoyang Zhang, Yuxin Li, Daijiao Liu, Yayue Deng, Donghang Wu, Jun Chen, Liang Zhao, Chengyuan Yao, Hexin Liu, Eng~Siong Chng, Xuerui Yang, Xiangyu Zhang, Daxin Jiang, and Gang Yu. 2025.
\newblock \href {https://doi.org/10.48550/arXiv.2511.15848} {Step-audio-r1 technical report}.
\newblock \emph{arXiv preprint arXiv:2511.15848}.

\bibitem[{Tomar(2006)}]{tomar2006converting}
Suramya Tomar. 2006.
\newblock Converting video formats with ffmpeg.
\newblock \emph{Linux journal}, 2006(146):10.

\bibitem[{Tong et~al.(2026)Tong, Li, Wang, Bi, Cai, Liu, He, and Hao}]{tong2026autagent}
Siqian Tong, Xuan Li, Yiwei Wang, Baolong Bi, Yujun Cai, Shenghua Liu, Yuchen He, and Chengpeng Hao. 2026.
\newblock \href {https://doi.org/10.48550/arXiv.2602.13685} {Autagent: A reinforcement learning framework for tool-augmented audio reasoning}.
\newblock \emph{arXiv preprint arXiv:2602.13685}.

\bibitem[{Wan et~al.(2026)Wan, Yang, Tian, Ye, Pasad, Fu, Goel, Hachiuma, Diao, Dhawan, Ghosh, Hirota, Chen, Valle, Hosseini~Asl, Chu, Watanabe, Wang, and Ginsburg}]{wan2026speechhands}
Zhen Wan, Chao-Han~Huck Yang, Jinchuan Tian, Hanrong Ye, Ankita Pasad, Szu-wei Fu, Arushi Goel, Ryo Hachiuma, Shizhe Diao, Kunal Dhawan, Sreyan Ghosh, Yusuke Hirota, Zhehuai Chen, Rafael Valle, Ehsan Hosseini~Asl, Chenhui Chu, Shinji Watanabe, Yu-Chiang~Frank Wang, and Boris Ginsburg. 2026.
\newblock \href {https://doi.org/10.48550/arXiv.2601.09413} {Speech-hands: A self-reflection voice agentic approach to speech recognition and audio reasoning with omni perception}.
\newblock \emph{arXiv preprint arXiv:2601.09413}.

\bibitem[{Wang et~al.(2023)Wang, Liang, Wang, Chen, Zhang, Xiang, Deng, and Qian}]{wang2023wespeaker}
Hongji Wang, Chengdong Liang, Shuai Wang, Zhengyang Chen, Binbin Zhang, Xu~Xiang, Yanlei Deng, and Yanmin Qian. 2023.
\newblock \href {https://doi.org/10.48550/arXiv.2210.17016} {Wespeaker: A research and production oriented speaker embedding learning toolkit}.
\newblock In \emph{Proceedings of the IEEE International Conference on Acoustics, Speech and Signal Processing}.

\bibitem[{Wang et~al.(2025)Wang, Sun, Lin, Wang, Pan, and Xie}]{wang2025msubench}
Shuai Wang, Zhaokai Sun, Zhennan Lin, Chengyou Wang, Zhou Pan, and Lei Xie. 2025.
\newblock \href {https://doi.org/10.48550/arXiv.2508.08155} {{MSU-Bench}: Towards understanding the conversational multi-talker scenarios}.
\newblock \emph{arXiv preprint arXiv:2508.08155}.

\bibitem[{Wijngaard et~al.(2025)Wijngaard, Formisano, Dumontier, and Jitsev}]{wijngaard2025audiotoolagent}
Gijs Wijngaard, Elia Formisano, Michel Dumontier, and Jenia Jitsev. 2025.
\newblock \href {https://doi.org/10.48550/arXiv.2510.02995} {Audiotoolagent: An agentic framework for audio-language models}.
\newblock \emph{arXiv preprint arXiv:2510.02995}.

\bibitem[{Wu et~al.(2025)Wu, Yan, Hu, Yi, Feng, Tian, Shen, Yu, Zhang, Li et~al.}]{wu2025step}
Boyong Wu, Chao Yan, Chen Hu, Cheng Yi, Chengli Feng, Fei Tian, Feiyu Shen, Gang Yu, Haoyang Zhang, Jingbei Li, and 1 others. 2025.
\newblock Step-audio 2 technical report.
\newblock \emph{arXiv preprint arXiv:2507.16632}.

\bibitem[{Xiong et~al.(2025)Xiong, Cai, Li, Yuan, and Wang}]{xiong2025thinkingwithsound}
Zhen Xiong, Yujun Cai, Zhecheng Li, Junsong Yuan, and Yiwei Wang. 2025.
\newblock \href {https://doi.org/10.48550/arXiv.2509.21749} {Thinking with sound: Audio chain-of-thought enables multimodal reasoning in large audio-language models}.
\newblock \emph{arXiv preprint arXiv:2509.21749}.

\bibitem[{Xu et~al.(2025{\natexlab{a}})Xu, Guo, Hu, Chu, Wang, He, Wang, Shi, He, Zhu et~al.}]{xu2025qwen3}
Jin Xu, Zhifang Guo, Hangrui Hu, Yunfei Chu, Xiong Wang, Jinzheng He, Yuxuan Wang, Xian Shi, Ting He, Xinfa Zhu, and 1 others. 2025{\natexlab{a}}.
\newblock Qwen3-omni technical report.
\newblock \emph{arXiv preprint arXiv:2509.17765}.

\bibitem[{Xu et~al.(2025{\natexlab{b}})Xu, Xie, Tang, and Hu}]{xu2025fireredasr}
Kai-Tuo Xu, Feng-Long Xie, Xu~Tang, and Yao Hu. 2025{\natexlab{b}}.
\newblock \href {https://doi.org/10.48550/arXiv.2501.14350} {Fireredasr: Open-source industrial-grade mandarin speech recognition models from encoder-decoder to llm integration}.
\newblock \emph{arXiv preprint arXiv:2501.14350}.

\bibitem[{Xu et~al.(2026)Xu, Jia, Huang, Chen, Li, Liu, Xie, Tang, and Hu}]{xu2026fireredasr2s}
Kaituo Xu, Yan Jia, Kai Huang, Junjie Chen, Wenpeng Li, Kun Liu, Feng-Long Xie, Xu~Tang, and Yao Hu. 2026.
\newblock \href {https://doi.org/10.48550/arXiv.2603.10420} {Fireredasr2s: A state-of-the-art industrial-grade all-in-one automatic speech recognition system}.
\newblock \emph{arXiv preprint arXiv:2603.10420}.

\bibitem[{Yang et~al.(2024)Yang, Xu, Liu, Chu, Jiang, Zhou, Leng, Lv, Zhao, Zhou, and Zhou}]{yang2024airbench}
Qian Yang, Jin Xu, Wenrui Liu, Yunfei Chu, Ziyue Jiang, Xiaohuan Zhou, Yichong Leng, Yuanjun Lv, Zhou Zhao, Chang Zhou, and Jingren Zhou. 2024.
\newblock \href {https://doi.org/10.48550/arXiv.2402.07729} {{AIR-Bench}: Benchmarking large audio-language models via generative comprehension}.
\newblock \emph{arXiv preprint arXiv:2402.07729}.

\bibitem[{Yao et~al.(2022)Yao, Zhao, Yu, Du, Shafran, Narasimhan, and Cao}]{yao2022react}
Shunyu Yao, Jeffrey Zhao, Dian Yu, Nan Du, Izhak Shafran, Karthik Narasimhan, and Yuan Cao. 2022.
\newblock \href {https://doi.org/10.48550/arXiv.2210.03629} {{ReAct}: Synergizing reasoning and acting in language models}.
\newblock \emph{arXiv preprint arXiv:2210.03629}.

\bibitem[{Zhang et~al.(2025)Zhang, Wang, Xue, Fang, Zhao, Ma, Ren, Liu, Guo, Zhuang et~al.}]{zhang2025mimo}
Dong Zhang, Gang Wang, Jinlong Xue, Kai Fang, Liang Zhao, Rui Ma, Shuhuai Ren, Shuo Liu, Tao Guo, Weiji Zhuang, and 1 others. 2025.
\newblock Mimo-audio: Audio language models are few-shot learners.
\newblock \emph{arXiv preprint arXiv:2512.23808}.

\end{thebibliography}

\appendix

\begin{table*}[t]
\scriptsize
\centering
\setlength{\tabcolsep}{3.5pt}
\resizebox{\textwidth}{!}{%
\begin{tabular}{llrrrrrr}
\toprule
\textbf{Layer} & \textbf{Sub-category} & \textbf{N} & \textbf{Qwen} & \textbf{AudioGenie} & \textbf{\ours{}} & \textbf{$\Delta$ vs. Qwen} & \textbf{$\Delta$ vs. AudioGenie} \\
\midrule
Cultural & Aesthetic Evaluation & 8 & 75.00 & 50.00 & 62.50 & -12.50 & +12.50 \\
Cultural & Culture of Speaker & 52 & 78.85 & 76.92 & 82.69 & +3.85 & +5.77 \\
Cultural & Imagination & 10 & 70.00 & 60.00 & 80.00 & +10.00 & +20.00 \\
Cultural & Professional Knowledge and Reasoning & 71 & 76.06 & 74.65 & 76.06 & +0.00 & +1.41 \\
Perception & Correlation Analysis & 50 & 80.00 & 74.00 & 88.00 & +8.00 & +14.00 \\
Perception & Counting and Statistics & 99 & 64.65 & 56.57 & 63.64 & -1.01 & +7.07 \\
Perception & Environmental Perception and Reasoning & 149 & 87.25 & 70.47 & 85.91 & -1.34 & +15.44 \\
Perception & Music Theory & 63 & 57.14 & 55.56 & 73.02 & +15.87 & +17.46 \\
Perception & Spatial Analysis & 15 & 73.33 & 73.33 & 60.00 & -13.33 & -13.33 \\
Perception & Temporal Analysis & 28 & 64.29 & 42.86 & 78.57 & +14.29 & +35.71 \\
Semantic & Content Analysis & 304 & 85.20 & 80.26 & 86.84 & +1.64 & +6.58 \\
Semantic & Emotion and Intention & 60 & 83.33 & 65.00 & 76.67 & -6.67 & +11.67 \\
Semantic & Speaker Analysis & 48 & 87.50 & 77.08 & 87.50 & +0.00 & +10.42 \\
Signal & Acoustic Quality Analysis & 18 & 77.78 & 55.56 & 66.67 & -11.11 & +11.11 \\
Signal & Anomaly Detection & 17 & 70.59 & 70.59 & 88.24 & +17.65 & +17.65 \\
Signal & Audio Difference Analysis & 8 & 62.50 & 50.00 & 37.50 & -25.00 & -12.50 \\
\bottomrule
\end{tabular}
}
\caption{Full MMAR sub-category accuracy comparison. Values are accuracies in percentage points. AudioGenie denotes AudioGenie-Reasoner instantiated with Qwen3.5-Omni and Qwen3.5 for matched-backbone comparison.}
\label{tab:mmar_subcategory_full}
\end{table*}

\section{MMAR Sub-Category Results}
\label{sec:mmar_subcategory_appendix}

Table~\ref{tab:mmar_subcategory_full} reports the MMAR sub-category comparison among Qwen3.5-Omni, matched-backbone AudioGenie-Reasoner, and \ours{} on the same 1,000-example evaluation split.

The sub-category results suggest that \ours{} helps most when a question can benefit from bounded, task-specific evidence rather than generic decomposition. The largest gains over Qwen3.5-Omni appear in Music Theory (+15.87), Temporal Analysis (+14.29), Anomaly Detection (+17.65), and Correlation Analysis (+8.00), which are categories where structured measurements, temporal localization, or transformed audio segments can provide concrete support for the frontend. In contrast, \ours{} does not consistently improve categories that rely mainly on holistic semantic judgment or perceptual interpretation: it is slightly worse on Spatial Analysis, Acoustic Quality Analysis, and Audio Difference Analysis. This mixed pattern shows that there is no universally optimal division of labor between external tools and frontend LALMs. The capability boundary is question dependent: tools may provide more reliable evidence for some subcategories, while direct frontend perception may remain stronger for others. As both LALMs and specialized tools improve, this boundary may also shift across model generations and task settings. Estimating when tool evidence is likely to help, and when it may distract from stronger frontend judgment, is therefore an important direction for audio agent research.

\section{System Behavior and Tool-Usage Statistics}
\label{sec:behavior_appendix}

This appendix expands on Section~\ref{sec:behavior_analysis} with full distributions over planner rounds, tool calls, tool-usage frequency, and targeted re-listening. Unless otherwise stated, all behavioral statistics are computed on the 999 successfully completed MMAR runs; one of the 1,000 evaluation examples failed due to an API content-safety rejection and is excluded from tool-use statistics.
\subsection{Planner-Round Distribution}

A planner round is one decision step in which the planner emits either a tool call, a re-listen request, an answer, or a give-up action. Across the 999 runs, the planner uses on average 2.68 rounds per question. 99.9\% of runs exit with an \textsc{answer} action; only one run exits with \textsc{fail} due to an API content-safety rejection. Table~\ref{tab:planner_rounds_hist} reports the full distribution: 31.1\% of questions exit in a single round (i.e., the planner emits \textsc{answer} immediately after the initial caption), and 93.3\% exit within five rounds.

\begin{table}[t]
\small
\centering
\begin{tabular}{lrr}
\toprule
\textbf{Rounds} & \textbf{Count} & \textbf{Share} \\
\midrule
1 & 311 & 31.1\% \\
2 & 255 & 25.5\% \\
3 & 164 & 16.4\% \\
4--5 & 203 & 20.3\% \\
6--10 & 59 & 5.9\% \\
$>$10 & 7 & 0.7\% \\
\bottomrule
\end{tabular}
\caption{Distribution of planner rounds per MMAR question.}
\label{tab:planner_rounds_hist}
\end{table}

\subsection{Tool-Usage Frequency}

Across 999 MMAR runs, \ours{} invokes 1{,}675 tools in total. Table~\ref{tab:top_tools} lists the ten most frequently used tools, which together account for 74.5\% of all invocations. The distribution is heavy-tailed: ASR with timestamps, audio-plot inspection, and time-range trimming dominate, consistent with the prevalence of speech, sound, and temporal-localization questions in MMAR. Of the remaining 25 tools each accounts for less than 3\% of calls, but each is used on at least one question, indicating that the long tail of the inventory provides bounded, occasionally decisive coverage rather than redundant capacity. We also observed 6 invocations of a non-existent tool name (\texttt{content}) across 4 questions; these are planner hallucinations rejected by the executor and represent a minor failure mode in tool selection.

\begin{table}[t]
\small
\centering
\resizebox{\columnwidth}{!}{%
\begin{tabular}{lrrr}
\toprule
\textbf{Tool} & \textbf{Calls} & \textbf{\% of calls} & \textbf{Q-share} \\
\midrule
\texttt{transcribe\_qwenasr\_with\_timestamps} & 227 & 14.9 & 18.9 \\
\texttt{inspect\_audio\_plots} & 206 & 13.6 & 11.1 \\
\texttt{trim\_audio} & 173 & 11.4 & 6.9 \\
\texttt{transcribe\_fireredasr\_with\_timestamps} & 113 & 7.4 & 11.3 \\
\texttt{analyze\_pitch} & 107 & 7.0 & 5.4 \\
\texttt{analyze\_onsets} & 85 & 5.6 & 7.5 \\
\texttt{transcribe\_whisperx\_with\_diarization} & 64 & 4.2 & 6.2 \\
\texttt{extract\_rms\_energy} & 58 & 3.8 & 3.7 \\
\texttt{analyze\_spectral\_features} & 55 & 3.6 & 3.8 \\
\texttt{transcribe\_qwenasr} & 46 & 3.0 & 3.7 \\
\bottomrule
\end{tabular}
}
\caption{Top-10 most invoked tools on MMAR. Q-share is the percentage of questions invoking the tool at least once.}
\label{tab:top_tools}
\end{table}

\subsection{Targeted Re-Listening}

Targeted re-listening, in which the planner sends a selected audio identifier and a focused prompt back to the frontend, is used sparingly: only 50 out of 999 MMAR questions (5.0\%) trigger at least one re-listen, for 64 re-listens in total. This pattern is consistent with the intended role of re-listening: it is reserved for cases where tool-side transformation or selection has changed what the frontend can perceive, rather than serving as a generic fallback.

\begin{figure*}[t]
  \centering
  \resizebox{\textwidth}{!}{\begin{tikzpicture}[
  font=\small,
  panel/.style={
    draw=black!35,
    rounded corners=2pt,
    line width=0.5pt,
    fill=black!2,
    inner sep=0pt
  },
  header/.style={
    font=\bfseries,
    anchor=north west,
    align=left
  },
  box/.style={
    draw=black!25,
    rounded corners=2pt,
    line width=0.45pt,
    fill=white,
    align=left,
    inner sep=5pt
  },
  badbox/.style={
    draw=yellow!55!black,
    rounded corners=2pt,
    line width=0.55pt,
    fill=yellow!22,
    align=left,
    inner sep=5pt
  },
  unsurebox/.style={
    draw=blue!50!black,
    rounded corners=2pt,
    line width=0.55pt,
    fill=blue!8,
    align=left,
    inner sep=5pt
  },
  toolbox/.style={
    draw=teal!55!black,
    rounded corners=2pt,
    line width=0.55pt,
    fill=teal!7,
    align=left,
    inner sep=5pt
  },
  answerbox/.style={
    draw=green!45!black,
    rounded corners=2pt,
    line width=0.65pt,
    fill=green!9,
    align=left,
    inner sep=5pt
  },
  arrow/.style={-{Latex[length=2.2mm,width=1.6mm]}, thick, draw=black!45}
]

\node[panel, minimum width=6.4cm, minimum height=8.65cm, anchor=north west] (directPanel) at (0,0) {};
\node[panel, minimum width=10.5cm, minimum height=8.65cm, anchor=north west] (agentPanel) at (6.75cm,0) {};

\node[header] at ([xshift=0.20cm,yshift=-0.25cm]directPanel.north west)
  {Qwen3.5-Omni};
\node[anchor=north west, font=\scriptsize, text=black!65, align=left]
  at ([xshift=0.20cm,yshift=-0.58cm]directPanel.north west)
  {Commits to an unsupported acoustic cue};

\node[box, text width=5.75cm, minimum height=1.42cm, anchor=north west] (directPred)
  at ([xshift=0.20cm,yshift=-1.05cm]directPanel.north west)
  {\textbf{Prediction}\\
   First segment = hot water\\
   Second segment = cold water\\
   \textcolor{red!70!black}{Incorrect}};

\node[badbox, text width=5.75cm, minimum height=1.58cm, anchor=north west] (directHallucination)
  at ([xshift=0.20cm,yshift=-2.85cm]directPanel.north west)
  {\textbf{Highlighted hallucinated cue}\\
   ``The first pouring segment ... \textbf{fizzing, crackling, or sizzling}''\\[2pt]
   ``The second ... \textbf{smoother and quieter}''};

\node[box, text width=5.75cm, minimum height=1.72cm, anchor=north west] (directIssue)
  at ([xshift=0.20cm,yshift=-5.05cm]directPanel.north west)
  {\textbf{Failure mode}\\
   The model gives a plausible physical explanation, but the claimed high-frequency cue is assigned to the wrong segment.};

\node[header] at ([xshift=0.20cm,yshift=-0.25cm]agentPanel.north west)
  {Audio-Mind workflow};
\node[anchor=north west, font=\scriptsize, text=black!65, align=left]
  at ([xshift=0.20cm,yshift=-0.58cm]agentPanel.north west)
  {Uncertainty is exposed, then resolved with tools and targeted re-listening};

\node[unsurebox, text width=9.85cm, minimum height=1.58cm, anchor=north west] (caption)
  at ([xshift=0.20cm,yshift=-1.05cm]agentPanel.north west)
  {\textbf{1. Question-oriented frontend caption}\\
   Locates Segment 1 at 3--8s and Segment 2 at 19--24s.\\
   States: ``\textbf{acoustic evidence is insufficient}'' and ``\textbf{no audible steam hiss}''.\\
   Forced hypothesis confidence: \textbf{0.3}.};

\node[toolbox, text width=9.85cm, minimum height=1.62cm, anchor=north west] (tools)
  at ([xshift=0.20cm,yshift=-3.18cm]agentPanel.north west)
  {\textbf{2. Evidence log}\\
   \texttt{trim\_audio}: creates \texttt{audio\_1} (first pour) and \texttt{audio\_2} (second pour).\\
   \texttt{analyze\_spectral\_features}: Segment 2 has higher centroid
   (5195 vs. 4013 Hz), rolloff (10844 vs. 8921 Hz), and flatness (0.011 vs. 0.004).};

\node[unsurebox, text width=9.85cm, minimum height=1.44cm, anchor=north west] (followup)
  at ([xshift=0.20cm,yshift=-5.35cm]agentPanel.north west)
  {\textbf{3. Targeted frontend follow-up}\\
   Re-listens to isolated segments with spectral evidence.\\
   States: ``The second audio segment contains ... \textbf{high-frequency hissing and sizzling}''.};

\node[answerbox, text width=9.85cm, minimum height=0.84cm, anchor=north west] (answer)
  at ([xshift=0.20cm,yshift=-7.45cm]agentPanel.north west)
  {\textbf{4. Final answer}\\
   \textbf{Second segment = hot water; first segment = cold water.}
   \hfill \textcolor{green!45!black}{Correct}};

\draw[arrow] (caption.south) -- (tools.north);
\draw[arrow] (tools.south) -- (followup.north);
\draw[arrow] (followup.south) -- (answer.north);

\end{tikzpicture}}
  \caption{Qualitative correction trace for a hot-versus-cold water example. The left panel shows the Qwen3.5-Omni response, where highlighted text marks the unsupported acoustic cue that leads to the wrong answer. The right panel shows the agent path: uncertainty is stated in the initial caption, tools isolate and measure both pouring segments, and targeted frontend follow-up uses this evidence to produce the correct answer.}
  \label{fig:case_study_hot_water}
\end{figure*}
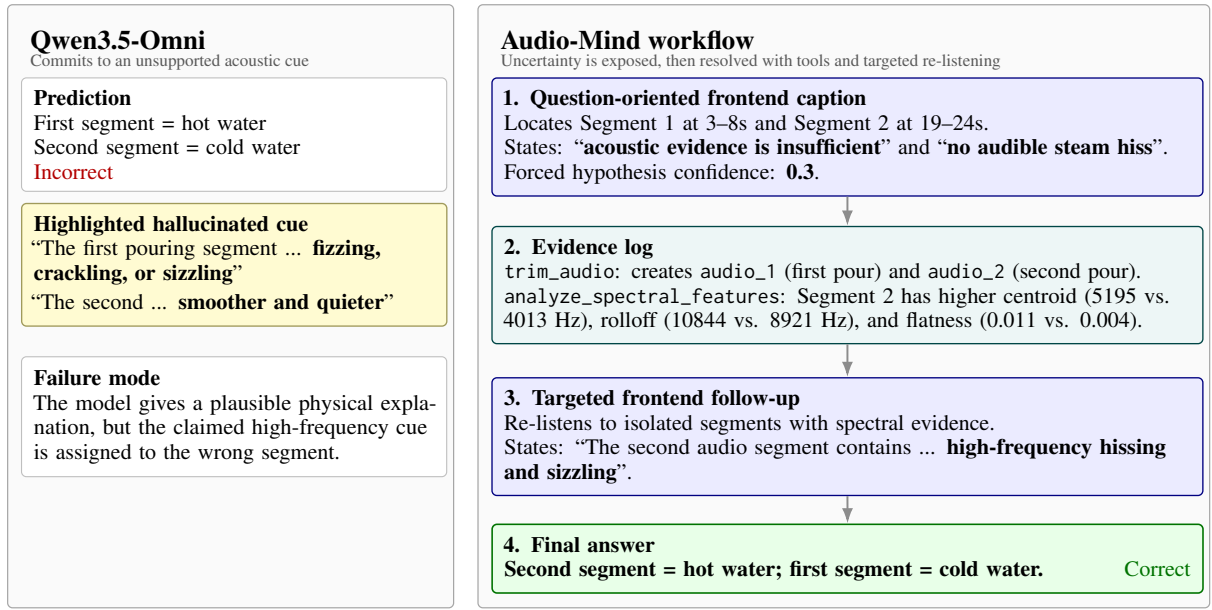

\subsection{Long-Tail \texorpdfstring{$>$}{>}10-Call Bucket}

In Section~\ref{sec:behavior_analysis}, the $>10$ tool-call bucket on MMAR contains only six questions and is too small to support a reliable interpretation. On these six questions, the direct Qwen3.5-Omni baseline achieves 50.0\% accuracy while \ours{} drops to 16.7\%. Possible explanations include long tool chains accumulating noisy intermediate evidence, planner instability on the hardest examples, and selection bias --- only the most uncertain questions reach this depth. Distinguishing these explanations would require more examples than this benchmark provides at the tail; we therefore treat the bucket as suggestive of a degradation regime worth investigating, but do not draw conclusions from it.

\section{Qualitative Case Study}
\label{sec:case_study}

We include one qualitative case to illustrate three behaviors that motivate the agent design. First, \textbf{an LALM may hallucinate a plausible acoustic rationale}: in an MMAR example asking which of two water-pouring segments is hot, Qwen3.5-Omni answers incorrectly by attributing a fizzing or sizzling cue to the first segment. Second, \textbf{the question-oriented frontend prompt helps expose uncertainty}: under our prompt, the same frontend localizes the two pours at approximately 3--8s and 19--24s, but marks the raw perceptual evidence as weak, notes that no clear steam hiss or ice clink is audible, and assigns only 0.3 confidence to its forced hypothesis. Third, \textbf{tools provide the missing evidence}: \ours{} isolates the two candidate segments, measures their spectral properties, and asks the frontend to re-listen with this comparative evidence. The second segment shows stronger high-frequency noise-like energy, with higher spectral centroid (5195 Hz vs. 4013 Hz), rolloff (10844 Hz vs. 8921 Hz), and flatness (0.011 vs. 0.004), leading targeted re-listening to reverse the initial forced option and select the correct answer. This example does not establish broad tool reliability by itself, but illustrates how \ours{} turns exposed uncertainty into focused evidence acquisition, as shown in Figure~\ref{fig:case_study_hot_water}.

\section{Reproducibility Details}
\label{sec:reproducibility_appendix}

The MMAR configuration uses \texttt{qwen3.5-omni-plus} for the audio-capable frontend and \texttt{qwen3.5-plus} for the text planner. The MSU-Bench configuration uses \texttt{Gemini 2.5} with thinking enabled as the audio-capable frontend and \texttt{Gemini 3.0 Flash} with thinking enabled as the text planner. We use temperature 0.05 for model calls and allow at most 15 planner decision steps per example. We use all pretrained models, benchmarks, and external tools under their respective licenses or terms of use, only for research evaluation and analysis, and do not redistribute third-party model weights, benchmark data, or tool artifacts. The prompt templates, tool inventory, and evaluation scripts are included with the submitted supplementary materials and will be released with the code.

\section{Tool Inventory}
\label{sec:tool_inventory_appendix}

Table~\ref{tab:tool_inventory_full} summarizes the tool inventory used in our experiments. We use the term ``tool'' broadly: some entries wrap specialized pretrained models, such as ASR, diarization, speaker verification, or chord recognition systems, while others are deterministic or library-based audio operations implemented with standard signal-processing utilities such as FFmpeg~\citep{tomar2006converting} and librosa~\citep{mcfee2015librosa}. The planner sees both types through the same bounded interface, but their expected outputs and reliability assumptions are specified separately in the tool-inventory configuration. Detailed input schemas and boundary instructions are maintained in the planner configuration.

\begin{table*}[t]
\scriptsize
\centering
\renewcommand{\arraystretch}{1.05}
\begin{tabularx}{\textwidth}{>{\raggedright\arraybackslash}p{0.28\textwidth}>{\raggedright\arraybackslash}X}
\toprule
\textbf{Tool} & \textbf{Description} \\
\midrule
\multicolumn{2}{c}{\emph{Metadata validation}} \\
\cmidrule(lr){1-2}
Audio stream statistics & Reports audio stream metadata and summary statistics. \\
Loading metadata & Reports duration, sample rate, and basic loading metadata. \\
\midrule
\multicolumn{2}{c}{\emph{Audio derivation}} \\
\cmidrule(lr){1-2}
FFT denoising & Creates a denoised audio file using FFT-domain filtering. \\
Wavelet denoising & Creates a denoised audio file using wavelet filtering. \\
Channel conversion & Converts audio to a target channel layout. \\
High-pass filtering & Attenuates low frequencies below a cutoff. \\
Low-pass filtering & Attenuates high frequencies above a cutoff. \\
Resampling & Converts audio to a target sample rate. \\
Harmonic--percussive separation & Estimates harmonic and percussive source components. \\
Time-range trimming & Cuts a selected time span into a derived clip. \\
\midrule
\multicolumn{2}{c}{\emph{Temporal segmentation}} \\
\cmidrule(lr){1-2}
Speech/music activity detection & Detects coarse speech, singing, and music regions. \\
FireRed speech VAD~\citep{xu2026fireredasr2s} & Detects likely speech activity regions. \\
Audio segmentation & Splits audio using fixed-duration or silence-based rules. \\
Silence detection & Detects silence intervals with threshold and duration settings. \\
Silero VAD & Detects likely speech regions with Silero VAD. \\
\midrule
\multicolumn{2}{c}{\emph{Speech and speaker processing}} \\
\cmidrule(lr){1-2}
DiariZen diarization~\citep{han2025efficient_diarization} & Estimates anonymous speaker-labeled time segments. \\
SortFormer diarization~\citep{park2025sortformer} & Estimates speaker-labeled segments with arrival-order tracking. \\
Speaker verification~\citep{wang2023wespeaker} & Compares two clips for same-speaker evidence. \\
FireRedASR~\citep{xu2025fireredasr,xu2026fireredasr2s} & Transcribes speech without word timestamps. \\
FireRedASR with timestamps~\citep{xu2026fireredasr2s} & Transcribes speech with rough timing information. \\
Qwen3-ASR~\citep{shi2026qwen3asr} & Transcribes speech with the Qwen ASR model. \\
Qwen3-ASR with timestamps~\citep{shi2026qwen3asr} & Transcribes speech with timestamp information. \\
WhisperX ASR~\citep{bain2023whisperx} & Transcribes speech with WhisperX. \\
WhisperX ASR with diarization~\citep{bain2023whisperx,bredin2023pyannote,plaquet2023powerset} & Produces transcription with diarization output. \\
\midrule
\multicolumn{2}{c}{\emph{Acoustic/music feature analysis}} \\
\cmidrule(lr){1-2}
Onset analysis & Detects onset times and onset-strength cues. \\
Pitch contour analysis & Estimates rough pitch contours and pitch statistics. \\
Tempo-CNN estimation~\citep{schreiber2018tempocnn} & Estimates two likely tempi and relative salience. \\
Spectral descriptors & Computes centroid, bandwidth, rolloff, contrast, and flatness. \\
Amplitude statistics & Computes detailed amplitude statistics. \\
Musical key estimation & Estimates a coarse musical key center. \\
Chroma extraction & Extracts pitch-class energy features over time. \\
MFCC extraction & Extracts spectral-envelope features. \\
RMS energy extraction & Extracts an energy-envelope proxy. \\
Large-vocabulary chord recognition~\citep{jiang2019largevocabularychord} & Recognizes chord timelines with extended chord vocabulary. \\
FFmpeg spectral statistics & Computes spectral statistics over audio. \\
Volume statistics & Computes mean and maximum volume statistics. \\
\midrule
\multicolumn{2}{c}{\emph{Signal visualization inspection}} \\
\cmidrule(lr){1-2}
Audio-plot inspection~\citep{bai2025qwen3} & Inspects waveforms, spectrograms, or other audio plots for bounded signal-level cues. \\
Processing-quality verification & Checks whether trimmed, filtered, separated, or denoised audio remains usable for downstream analysis. \\
\bottomrule
\end{tabularx}
\caption{Full planner-facing tool inventory used by \ours{}. Each category header groups tools with a shared operational role; detailed tool schemas and boundary instructions are specified in the planner inventory configuration.}
\label{tab:tool_inventory_full}
\end{table*}

\section{Generative AI Use Disclosure}

The authors used AI writing assistants to support language polishing, phrasing refinement, and checklist preparation. All scientific ideas, system design, experiments, analyses, and final manuscript decisions were made and verified by the authors.

\end{document}